\documentclass[12pt]{article}
\usepackage{latexsym,amsmath,amssymb,theorem,epsfig,epsf,rotate}
\def\f{\phi}
\def\z{\psi}
\begin{document}

\begin{titlepage}

\title{
   \hfill{\normalsize  UPR1130-T} \\[1em]
   {\LARGE The Dynamics of Small Instanton Phase Transitions}
\author{Alexander Borisov$^{1}$, Evgeny I.~Buchbinder$^{2}$, Burt A.~Ovrut$^{1}$
        \footnote{
        borisov@sas.upenn.edu;
        evgeny@ias.edu;
        ovrut@physics.upenn.edu
             }\\[0.5cm]
   {\normalsize $^1$ Department of Physics and Astronomy, University of
Pennsylvania,}\\[-0.4em]
   {\normalsize Philadelphia, PA 19104}\\
   {\normalsize $^2$ School of Natural Sciences, Institute for Advanced Study} \\[-0.4em]
{\normalsize Einstein Drive, Princeton, NJ 08540}\\
}
}
\date{}

\maketitle

\begin{abstract}

The small instanton transition of a five-brane colliding with one end of the ${\bf S \rm}^{1}/{\bf Z \rm}_{2}$ interval in heterotic $M$-theory is discussed, with emphasis on the transition moduli, their potential function and the associated non-perturbative superpotential. Using numerical methods, the equations of motion of these moduli coupled to an expanding Friedmann-Robertson-Walker spacetime are solved including non-perturbative interactions. It is shown that the five-brane collides with the end of the interval at a small instanton. However, the moduli then continue to evolve to an isolated minimum of the potential, where they are trapped by gravitational damping.
The torsion free sheaf at the small instanton is ``smoothed out'' into a vector bundle at the isolated minimum, thus dynamically completing the small instanton phase transition. Radiative damping at the origin of moduli space is discussed and shown to be insufficient to trap the moduli at the small instanton point.

\end{abstract}

\thispagestyle{empty}

\end{titlepage}

\section{Introduction:}

In~\cite{1}, Witten studied $SO(32)$ heterotic string theory on ${\bf R \rm}^6 \times
{\bf K3 \rm}$ with an $SO(N)$, $4 \leq N \leq 32$ gauge instanton supported
on $\bf K3 \rm$ with instanton number $24$. The effective $\bf R \rm^{6}$
theory manifests $N=2$ supersymmetry and has gauge symmetry $SO(32-N)$.
The boundary of instanton moduli space contains singular points,
at each of which
one unit of instanton number is concentrated. These are
called ``small'' instantons and the local moduli space ${\cal{M}}'$ was
shown to be $4N-12$ dimensional. It was demonstrated in~\cite{1} that ${\cal{M}}'$
can be described as the flat directions of a potential energy function.
This is constructed from $N$ hypermultiplets, each transforming as a
fundamental $N$ representation of $SO(N)$ and as a doublet under an
additional $SU(2)$ gauge symmetry. At the origin of field space, all
of these fields become massless and the $SU(2)$ symmetry
becomes exact. Away from this point, however, the $SU(2)$ symmetry is
spontaneously broken and only the $4N-12$ moduli of ${\cal{M}'}$
remain massless. One concludes that as one approaches a small instanton
point in moduli space, the theory develops extra (massless) species
and an enhanced symmetry. Such points are often referred to as
being $ESP$.

Small instantons also occur in $M$-theory on ${\bf R \rm}^{4} \times X
\times {\bf S \rm}^{1}/ {\bf Z \rm}_{2}$, where $X$ is a
Calabi-Yau threefold~\cite{2,2A}.
As first discussed in~\cite{2}, to be consistent these vacua must have,
supported on the Calabi-Yau manifold at each end of the
${\bf S \rm}^{1}/{\bf Z \rm}_{2}$ interval, a gauge instanton indexed
by a subgroup of $E_{8}$. One can, for simplicity, assume a trivial
instanton at one end of the interval. The non-trivial instanton on the other
end is a connection on a holomorphic vector bundle $V$ over $X$ with structure
group $G \subset E_{8}$. Quantum consistency requires that the instanton
number, here generalized to the 2nd Chern class $c_{2}(V)$, must satisfy
the constraint $c_{2}(V)=c_{2}(TX)-{\cal{W}}$, where $TX$ is the tangent
bundle of the Calabi-Yau threefold and ${\cal{W}}$ is an effective class corresponding
to M-five-branes in the ${\bf S \rm}^{1}/ {\bf Z \rm}_{2}$ ``bulk'' space.
The effective ${\bf R \rm}^{4}$ theory manifests $N=1$ supersymmetry.
Its gauge symmetry depends on the choice of structure group $G$. In a
series of publications~\cite{3,4}, holomorphic vector bundles with $G=SU(N)$ were
constructed. We will refer to these when specificity is required. For
example, when $G=SU(3)$, $SU(4)$ and $SU(5)$, the effective ${\bf R \rm}^{4}$
gauge symmetry is $E_{6}$, $Spin(10)$ and $SU(5)$ respectively.

The boundary of instanton moduli space contains singular points where a
portion of the vector bundle becomes concentrated as a torsion free sheaf over
an isolated holomorphic curve with effective class ${\cal{C}}'$. These are the heterotic
$M$-theory analogs of small instantons. Here, however, the local moduli spaces ${\cal{M}}'$
are more complicated, changing with the structure of the Calabi-Yau threefold, the
holomorphic vector bundle and the choice of ${\cal{C}'}$. The moduli spaces of small instantons
on elliptic Calabi-Yau threefolds over Hirzebruch bases
$B={\bf F \rm}_{r}$ with vector bundles constructed from spectral covers were
analyzed in several publications~\cite{5}. For example, consider $B={\bf F \rm}_{1}$, $G=SU(3)$
and ${\cal{C}'}={\cal{S}}$, with ${\cal{S}}$ a section of
${\bf F \rm}_{1}$.  Then, it was shown that
$dim{\cal{M}}'= 3(b-a)-6$, where $a,b$ are integers satisfying the constraints $a > 6$ and $b > a+3$.
It is expected that any ${\cal{M}}'$ can again be described as the flat
directions of a potential energy function. This would be constructed from chiral
supermultiplets transforming in some representation $R$ of $G$ and, possibly, as a representation $r$ of
an additional gauge symmetry $H$. At the origin of field space, all
of these fields become massless and the $H$ symmetry
becomes exact. Away from this point, the $H$ symmetry is
spontaneously broken and only the moduli of ${\cal{M}'}$
remain massless. Therefore, as one approaches a small instanton
in moduli space, the theory develops extra (massless) species and, possibly, an enhanced symmetry.
That is, small instantons in heterotic $M$-theory are $ESP$. Unfortunately, such potential functions have
not been explicitly constructed for most moduli spaces. Limited simple examples indicate that
such potentials exist, but a general proof is still lacking. Be that as it may, in this paper we will
assume that small instanton moduli spaces ${\cal{M}'}$ in heterotic $M$-theory can be so described.

In heterotic $M$-theory, strings wrapping a holomorphic curve in the Calabi-Yau threefold will generate a
non-perturbative superpotential for both geometric and vector bundle moduli~\cite{6,7}.
As first discussed in~\cite{8}, the superpotential created
by a such a ``worldsheet instanton'' has a specific form; namely, an exponential of
geometric and curve moduli multiplied by a ``Pfaffian''. The Pfaffian is a function of the determinant of the
Dirac operator twisted by the vector bundle restricted to the curve. It is generically true~\cite{9} that
the Pfaffian evaluated at any holomorphic curve is a homogeneous polynomial of the
moduli of the restricted bundle. This has important implications for small instantons.
It was shown in~\cite{9,10} that the moduli of the vector bundle restricted to the small instanton
curve ${\cal{C}}'$ are precisely those which parameterize ${\cal{M}}'$. Hence, strings wrapped on
${\cal{C}}'$ generate a superpotential on ${\cal{M}}'$ which ``lifts''some, or all, of the flat
directions. Furthermore, new local minima may now occur substantially away from the $ESP$ at the
zero of field space. Within the context of $B={\bf F \rm}_{r}$, $G=SU(3)$ and ${\cal{C}}'={\cal{S}}$,
Pfaffians were explicitly computed in~\cite{9,10}. For example, for $r=1$, $a>5 $ and $b=a+6$,
$dim{\cal{M}}'=3(6)-6=12$ and the Pfaffian was shown to be an explicit homogeneous polynomial
of degree $5$ in $7$ of the $12$ moduli of ${\cal{M}}'$.

At a small instanton in heterotic $M$-theory, the vector bundle $V$ degenerates into the ``union'' of a new holomorphic
vector bundle $V'$ over $X$ and a sheaf over the curve ${\cal{C}}'$. Because of the existence of the
${\bf S \rm}^{1}/ {\bf Z \rm}_{2}$ interval, the sheaf can ``detach'' from $X$,
leaving only $V'$ behind, and become a component of a new M-five-brane curve in the bulk space.
That is, ${\cal{C}}' \subset {\cal{W}}'$. The physical interpretation of this is that the
small instanton ``emits'' a five-brane into the bulk space. Of course, this can occur in reverse, a component ${\cal{C}}' \subset {\cal{W}}'$ of a bulk space five-brane being ``absorbed''
into $V'$ via a small instanton to become $V$. Such a process, either absorbing or emitting a
bulk five-brane, is called a small instanton phase transition and was defined and studied in~\cite{11}. It was shown that there are two types of
small instanton transitions,  the first in which the structure group of the vector bundle is unchanged while the number of chiral zero modes is altered
 and a second where the reverse is true, the  structure group changes while the number of chiral zero modes remains fixed.
For an elliptically fibered Calabi-Yau threefold $X$, the first type of transition occurs when ${\cal{C}}'$ is the pull-back of a curve in the base $B$. This is analogous to the heterotic $SO(32)$ case.
The second type of transition takes place when ${\cal{C}}'$ is proportional to the elliptic fiber $F$.
In this paper, we will consider only those small instanton transition for which the structure group is unchanged.

An important aspect of any small instanton transition is that the
$M$-five-brane wrapped on ${\cal{C}}'$ in the bulk space itself
has moduli. The most conspicuous of these is the translation
modulus, the real part of which describes the separation of the
five-brane from the end of the ${\bf S \rm}^{1} / {\bf Z
\rm}_{2}$ interval and whose imaginary part is an axion. For a
Calabi-Yau threefold $X$ elliptically fibered over any base $B$,
the complete moduli space ${\cal{M}}_{{\cal{C}}'}$ of an
$M$-five-brane wrapped on ${\cal{C}}'$, both its dimension and
geometry, was computed in~\cite{12}. This space depends strongly
on the curve ${\cal{C}}'$ and typically contains many moduli in
addition to the translation modulus. Combining this with the
discussion above, we see that in heterotic $M$-theory the full
moduli space of a small instanton phase transition is ${\cal{M}}'
\cup  {\cal{M}}_{{\cal{C}}'}$, where ${\cal{M}}'$ is the instanton
phase and ${\cal{M}}_{{\cal{C}}'}$ the five-brane phase.
Generalizing the above discussion, we assume that the complete
phase transition can be described by a potential energy function
constructed from chiral supermultiplets in some representation
$R$ of $G$ and, possibly, in a representation $r$ of an
additional gauge symmetry $H$. The origin of field space is an
$ESP$. Away from this point, however, the $H$ symmetry is
spontaneously broken and only the moduli parameterizing
${\cal{M}}' \cup {\cal{M}}_{{\cal{C}}'}$ remain massless.

The existence of small instanton phase transitions, from the point of view of
holomorphic vector bundles and topology, was demonstrated in~\cite{11}. In the present  paper, by studying the motion of heterotic $M$-theory vacua in moduli space coupled to an expanding Friedmann-Robinson-Walker (FRW) spacetime, we will show that these phase transition can also take place dynamically. Specifically, we will show that an $M$-five-brane in the bulk can dynamically evolve through the ESP, the point in moduli space where the five-brane collides with the end of the ${\bf S \rm}^{1} / {\bf Z \rm}_{2}$ interval,
continuing until the vacuum comes to rest at a local minimum of the non-perturbative superpotential. At this point, which is substantially distant from the ESP, the sheaf
over ${\cal{C}}'$ has combined with $V'$ to form a smooth vector bundle $V$, completing the small instanton phase transition. The dynamical system is trapped at this local minimum by the Hubble friction of the FRW space. We ignore the effect of moduli trapping due to radiation into massless fields at the ESP~\cite{13}, but will argue that this is insufficient to alter our results. Our results are consistent with those found, in different contexts, 
in~\cite{16,13,14,17,18}. These papers
ignore any non-perturbative potentials that might arise and discuss flat moduli space. They find that the moduli are trapped, both by Hubble friction and quantum radiation, exactly at the ESP, which, they go on to argue, gives it a special physical meaning. We note that were we to turn off the non-perturbative superpotential in heterotic $M$-theory, one would arrive at precisely the same conclusion. However, turning on this potential completely alters this result and allows the completion of the small instanton phase transition.

Throughout this paper, for notational and computational  simplicity,
we will take all fields to be dimensionless and choose units
in which $M_{P}=1$.
\newpage

\section{Moduli Trapping without a Non-Perturbative Superpotential:}

In this section, we consider the dynamics of small instanton
transitions ignoring the non-perturbative superpotential
generated by worldsheet instantons. In this case, the moduli space
${\cal{M}}' \cup {\cal{M}}_{{\cal{C}}'}$ is the flat locus of the
potential. As discussed above, the dimension and geometry of this
moduli space is completely understood, at least for a wide range
of base spaces $B$ and small instanton curves ${\cal{C}}'$.
However, ${\cal{M}}'$ and ${\cal{M}}_{{\cal{C}}'}$ are both,
generically, multi-dimensional. For example, the small instanton
discussed in the introduction had ${\cal{M}}'$ parameterized by
$12$ complex fields. Unfortunately, a detailed calculation of
dynamics in a multi-dimensional moduli space is not only
unenlightening in its complexity, but is far beyond our
computational power. For this reason, we will restrict our
discussion to a simplified moduli space, one in which both
${\cal{M}}'$ and ${\cal{M}}_{{\cal{C}}'}$ are taken to be
one-dimensional. We denote the complex modulus of ${\cal{M}}'$,
the instanton phase, by $\psi$. Physically, this modulus describes
the ``smoothing out'' of the small instanton sheaf into the smooth
vector bundle $V$. The complex modulus of
${\cal{M}}_{{\cal{C}}'}$, the five-brane phase, is written as
$\phi$ and represents the translation modulus of the five-brane.
We want to emphasize that the literature describing transitions in
different contexts~\cite{16,13,14,17,18} all employ very similar
simplifications.

\subsection*{Without Gravitation}

We begin by considering the dynamics of the moduli $\phi$ and
$\psi$ in static flat spacetime; that is, we turn off their
interaction with gravity. The simplest $N=1$ supersymmetric
self-interactions that have precisely a one-dimension instanton
phase and a one-dimension five-brane phase are described by the
superpotential
\begin{equation}
W=\lambda \phi^{2} \psi^{2}.
\label{1}
\end{equation}
Recall that both $\phi$ and $\psi$ are taken to be dimensionless and, hence, $\lambda$ has mass dimensions $3$ in Planck units. Furthermore, we choose
the Planck mass $M_{P}$ to be unity.
The potential function describing the small instanton phase transition is then given by
\begin{equation}
{\cal{V}}=|\partial_{\phi} W|^{2} + |\partial_{\psi} W|^{2}= 4 {\lambda}^{2} |\phi|^{2} |\psi|^{2}(|\phi|^{2}
+ |\psi|^{2}).
\label{2}
\end{equation}
The instanton phase is the one-dimensional complex submanifold
defined by $\phi=0$, whereas the five-brane phase is the
one-dimensional space given by $\psi=0$. The dynamical
calculation is carried out in the four-dimensional real space
parameterized by $Re \phi$, $Im \phi$, $Re \psi$ and $Im \psi$.
Without loss of generality, one can take $\lambda=1$.
Other values of $\lambda$ lead to
the same conclusions. 
In this paper, we will assume that 
the Calabi-Yau radius and the interval length are chosen, in units where
$M_{P}=1$, to be  in
the phenomenologically acceptable range
\begin{equation}
R_{CY} \sim (10^{-3})^{-1}, \quad \pi \rho \sim (10^{-4})^{-1}.
\label{13}
\end{equation}
This implies that in order to achieve the correct phenomenological
values for the Newton and gauge coupling parameters, the dimensionless volume and 
interval moduli should be stabilized at (or be slowly rolling near) the values of order one. 
In this paper, we will assume that this is the case.
This limits $\phi$ to be less
than or of the order of unity. For reasons to become clear in the
next subsection, we will choose both $\phi$ and $\psi$ moduli to have values much
smaller than one.

Let us compute the dynamics with the following initial conditions. Take the initial moduli to be
\begin{equation}
\phi_{0}= (7.4 \times 10^{-2}) + {\bf i \rm} (-4 \times 10^{-3}), \quad  \psi_{0}= (0) + {\bf i \rm} (3 \times 10^{-3})
\label{3}
\end{equation}
and their initial velocities as
\begin{equation}
{\dot{\phi}}_{0}= (-3.9 \times 10^{-7}) + {\bf i \rm} (-5.7 \times 10^{-7}), \quad  {\dot{\psi}_{0}}= (7.9
\times 10^{-6}) + {\bf i \rm} (-6.2 \times 10^{-6}).
\label{4}
\end{equation}
These values represent the following physical situation.  First,
there is a five-brane in the bulk space substantially separated
from the end of the ${\bf S \rm}^{1}/{\bf Z \rm}_{2}$ interval
with a much smaller displacement in the axionic direction. A
similar very small displacement occurs in the vector bundle
modulus. The five-brane has a small negative velocity in the
$Re\phi$ direction which will cause it to collide with the end
of the interval and a similar velocity in the axion direction.
Substantially larger velocities are given to both $Re\psi$ and
$Im\psi$ which enable a possible transition from the five-brane
branch to the instanton branch. The equations of motion are solved
numerically on MATHEMATICA, with the results for $Re\phi$ and
$Im\psi$ presented in Figures 1 and 2.
\begin{figure}[htb]
\center{\input{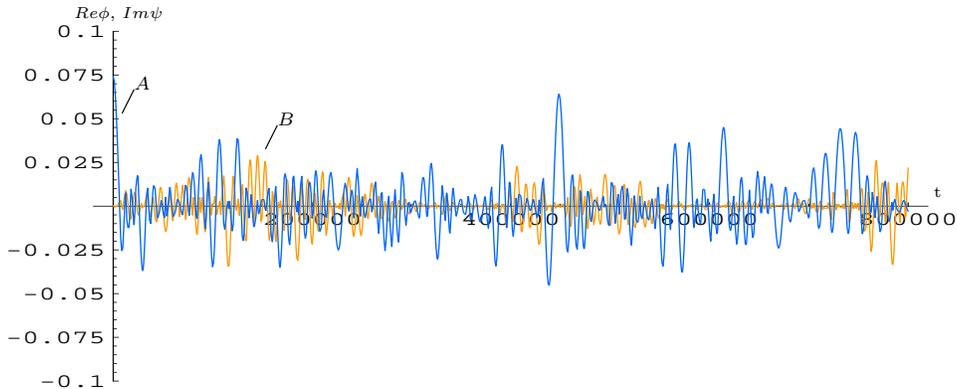}}
\caption{\label{pic11}{\footnotesize No gravitation and no non-perturbative superpotential:
curves (A) and (B) represent the dynamics of $Re\phi$ and $Im\psi$ respectively.}}
\end{figure}
Curve (A) in Figure 1 shows that the five-brane collides with
the end of the interval but then continues to oscillate around that point.
Similarly, curve (B) shows that the collision initiates a transition in the vector bundle modulus
direction which, however, quickly terminates. The resulting motion is an
oscillation around the small instanton in the vector bundle direction as well.
Figure 2 shows the overall motion in moduli space near the small instanton at the origin of field space.
Note that although the amplitude of any individual real modulus increases and decreases quasi-periodically,
the overall amplitude remains constant in time, as is clearly indicated in Figure 2. The dynamics of the
remaining two real moduli, $Im\phi$ and $Re\psi$, is similar, so we will not display them.

\begin{figure}[htb]
\center{\input{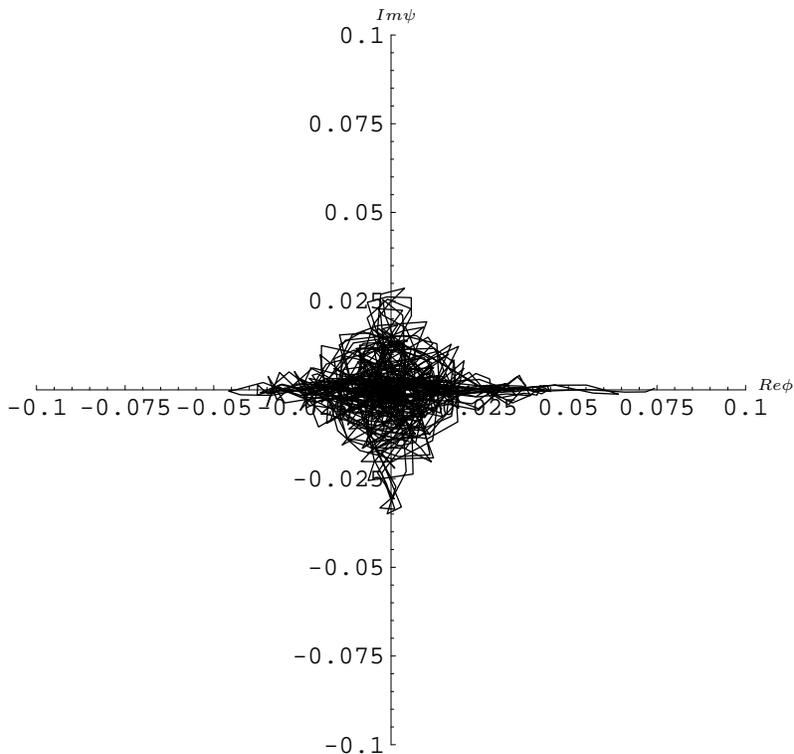}}
\caption{\label{pic12}{\footnotesize No gravitation and no non-perturbative superpotential:
the combined motion of $Re\phi$ and $Im\psi$ near the origin of moduli space.}}
\end{figure}
These results are easy to explain. First of all, in both the
five-brane and instanton branches, the potential energy function
has walls which become progressively steeper as one moves away
from the origin of field space. Hence, the moving moduli see a
``concave'' potential that always reflects them back toward the
origin. This accounts for the oscillation around the small
instanton at the origin of field space. Secondly, the lack of any
damping, be it Hubble friction or any other mechanism, means that
the overall oscillation amplitude persists. The final conclusion
is that the phase transition is never completed. Instead the
system oscillates endlessly around the small instanton. Note that
these results are consistent with those found in~\cite{16,13,14,17,18}
when analyzing undamped motion in different contexts.

\subsection*{With Gravitation}

Let us now consider the moduli $\phi$ and $\psi$ interacting with gravity. In this case, the dynamics is described
by $N=1$ chiral multiplets coupled to supergravity and one must specify both a superpotential $W$ and
a Kahler potential $K$. The superpotential remains the same, namely
\begin{equation}
W=\lambda \phi^{2} \psi^{2}.
\label{5}
\end{equation}
In general, the Kahler potential of moduli in heterotic $M$-theory is unknown. However,
near the origin of field space, that is, when the magnitudes of the moduli are much smaller than unity, this
can be taken to be ``flat''. Therefore, as long as we restrict $\phi$ and $\psi$ describing
our small instanton phase transition to be small, the Kahler potential is given by
\begin{equation}
K=|\phi|^{2}+|\psi|^{2}.
\label{6}
\end{equation}
This is one reason why we chose the initial conditions for $\phi$ and $\psi$ in the previous subsection
to be much less than one. The potential function describing the small instanton phase transition is now
\begin{equation}
{\cal{V}}=e^{K}
(|D_{\phi}|^{2}+|D_{\psi}|^{2}-3|W|^{2}),
\label{7}
\end{equation}
where $W$ and $K$ are given in~(\ref{5}) and~(\ref{6})
respectively and $D$ is the Kahler covariant derivative. This is
a rather complicated function of the moduli. However, for $\phi$
and $\psi$ much smaller than unity, one need only consider the
lowest order interactions. All terms that are higher order in
$M_{P}^{-1}$ can safely be ignored. The result is that one can
analyze the dynamics with the same potential as in the
non-gravitational case, that is,
\begin{equation}
{\cal{V}}=4{\lambda}^{2}|\phi|^{2}|\psi|^{2}(|\phi|^{2} + |\psi|^{2}).
\label{8}
\end{equation}
As previously, we will choose $\lambda=1$ and only consider the region of moduli space where $\phi$ and
$\psi$ are much smaller than one. Additionally, one must give an ansatz for
the spacetime geometry. We will assume a metric of the form
\begin{equation}
ds^{2}=-dt^{2}+a(t)^{2} {\vec{dx}}^{2}
\label{9}
\end{equation}
corresponding to a spatially flat manifold of the FRW type.
\begin{figure}[hb]
\center{\input{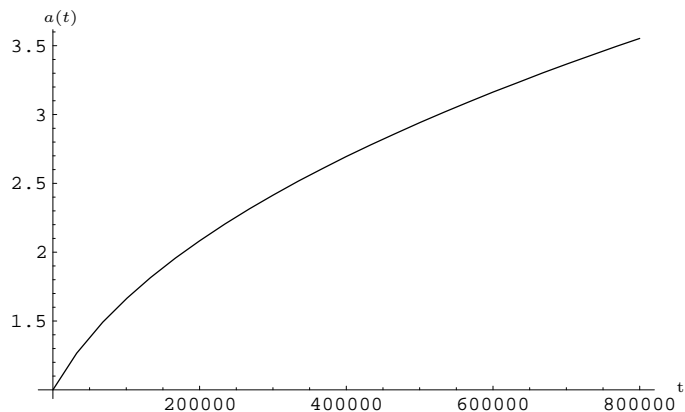}}
\caption{\label{pic13}{\footnotesize Graph of the FRW expansion
parameter $a(t)$.}}
\end{figure}
Furthermore, we will  only consider metrics that are
monotonically expanding. There are, in total, five equations of
motion that must be solved simultaneously. To begin with, there
is a single Einstein equation for the scale parameter $a$.
Additionally, there are the four equations of motion for the real
and imaginary parts of the moduli $\phi$ and $\psi$. Despite the
fact that the potential in~(\ref{8}) is independent of
gravitationally induced interactions, non-trivial gravity effects
do enter each equation as a Hubble damping term.

Let us compute the dynamics with the following initial
conditions. Take the initial value of the scale parameter to be
\begin{equation}
a_{0}=1.
\label{10}
\end{equation}
It is unnecessary to fix an initial radial velocity. We will
continue to choose the initial conditions for the $\phi$ and
$\psi$ moduli and their velocities to be those given in~(\ref{3})
and~(\ref{4}) respectively.
These values represent exactly the same physical situation as in
the previous subsection. Here, however, the moduli evolve with
respect to an expanding FRW background spacetime. The equations
of motion are solved numerically on MATHEMATICA. The result for
the scale parameter $a$ is shown in Figure 3, while those for
$Re\phi$ and $Im\psi$ are presented in Figures 4 and 5. Figure 3
shows that $a$ increases monotonically from its initial value,
indicating an expanding FRW universe.

\begin{figure}[htb]
\center{\input{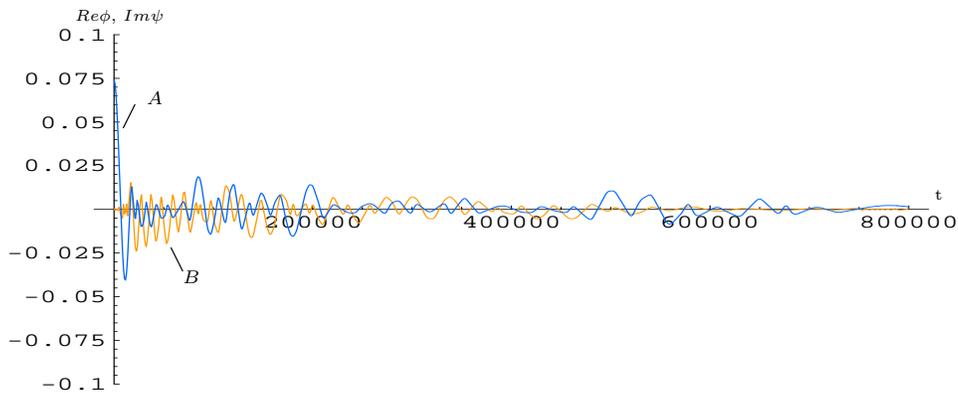}}
\caption{\label{pic14}{\footnotesize FRW spacetime but no
non-perturbative superpotential: curves (A) and (B) represent the
dynamics of $Re\phi$ and $Im\psi$ respectively.}}
\end{figure}

From curve (A) in Figure 4, one sees that the five-brane collides
with the end of the interval and then continues to oscillate
around that point. However, the average amplitude and frequency
of the oscillation are now strongly damped in time, with $Re\phi$
converging toward zero. Similarly, curve (B) in Figure 4 shows
that the collision initiates a transition into the vector bundle
modulus direction which then terminates, reverses and begins to
oscillates around $Im\psi=0$. Once again, the average amplitude
and frequency of the oscillation in the vector bundle modulus
direction are damped in time and $Im\psi$ converges to zero.
\begin{figure}[htb]
\center{\input{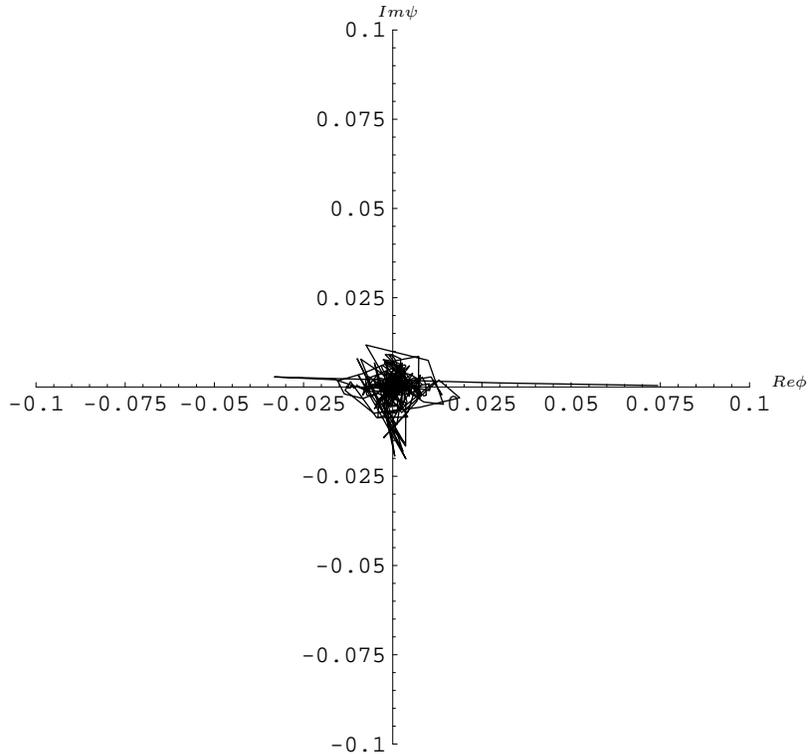}}
\caption{\label{pic15}{\footnotesize Gravitation but no
non-perturbative superpotential: the combined motion of $Re\phi$
and $Im\psi$ near the origin of moduli space.}}
\end{figure}
Figure 5 shows the overall motion in moduli space near the small
instanton at the origin of field space. Note that the overall
amplitude decreases in time. The dynamics of the remaining real
moduli, $Im\phi$ and $Re\psi$, is similar, so we do not display
them.

The physical explanation of these results is identical to that
given in the previous subsection, with one important exception.
The expansion of the universe produces ``friction'' in the motion
of the moduli. This causes their oscillation around the origin of
field space to have ever diminishing amplitude and frequency. The
physical conclusion is that after collision and a period of
damped oscillation, the five-brane becomes fixed at the end of
the interval. The associated bundle data is the ``union'' of  $V'$ and a torsion free sheaf over the curve ${\cal{C}}'$. That is, the system
approaches the small instanton $ESP$ and the phase transition to
a smooth bundle $V$ is never completed. Note that these results
are completely consistent with those found in~\cite{16,13,14,17,18} when
analyzing damped motion in different contexts.

%%%%%%%%%%%%%%%%%%%%%%%%%%%%%%%%%%%%%%%%%%%%%%%%%%%%%%%%%%%%%%%%%%%%%%%%%%%%%%%%%%%%

\section{Moduli Trapping with a Non-Perturbative Superpotential:}

%%%%%%%%%%%%%%%%%%%%%%%%%%%%%%%%%%%%%%%%%%%%%%%%%%%%%%%%%%%%%%%%%%%%%%%%%%%%%%%

Having established our preliminary results, we now turn to the
main discussion in this paper; namely, the dynamics of small
instanton transitions in the presence of a non-perturbative
superpotential generated by worldsheet instantons. In this case,
the moduli space ${\cal{M}}_{{\cal{C}}'}$ of the curve remains
flat. However, the potential for the vector bundle moduli is now
substantially altered, with the result that, generically, the flat
moduli space ${\cal{M}}'$ is replaced by one or more isolated
local minima. To proceed, we must first be more explicit about
the form of the non-perturbative interactions.

\subsection*{Heterotic Non-Perturbative Superpotentials}

It was shown in~\cite{8} that the superpotential generated by a string wrapped on an isolated curve
${{\cal{C}}'}$ has the form
\begin{equation}
W_{NP} \propto Pfaff(\psi_{a}) e^{-\tau \phi},
\label{11}
\end{equation}
where $\psi_{a}$ are a subset of vector bundle moduli and we have assumed, for simplicity, that the moduli space of
the five-brane wrapped on ${\cal{C}}'$ is one-dimensional and parameterized by the dimensionless translation modulus $\phi$.
The coefficient $\tau$ is then dimensionless and given by
\begin{equation}
\tau= \frac{1}{2} (2 \pi)^{1/3} \left( \frac{1}{ 2 \kappa^{2}_{11}} \right)^{1/3} \pi \rho v ,
\label{12}
\end{equation}
where $\rho$ is the size of the ${\bf S \rm}^{1}/{\bf Z \rm}_{2}$ interval, $v$ is the area of ${\cal{C}}'$
and $\kappa_{11}$ is the eleven-dimensional Planck scale. 
When the Calabi-Yau radius and the interval length are chosen, in units where
$M_{P}=1$, to be in
the phenomenologically acceptable range eq.~\eqref{13},
$\tau$ is found~\cite{15} to be of order $250$. 
Let us recall that in our notation the position of the five-brane is dimensionless
and bounded by the length of the interval which is of order one. 

As mentioned in the introduction, $Pfaff(\psi_{a})$ is the Pfaffian of the Dirac operator twisted by the vector bundle $V$ restricted to the curve ${\cal{C}}'$. It is found to be a homogeneous polynomial of the  moduli $\psi_{a}$ of the restricted bundle. In general, the Pfaffian is difficult to compute. However, it has been explicitly
calculated in~\cite{9,10} for
vector bundles with structure group $SU(N)$ restricted to
an isolated curve ${\cal{C}}'$ in an elliptic Calabi-Yau threefold with base
$B={\bf F \rm}_{r}$. For example, choosing $r=1$, a specific $SU(3)$ vector bundle and ${\cal{C}}'={\cal{S}}$, where ${\cal{S}}$ is an isolated curve in ${\bf F \rm}_{1}$, it was shown that the Pfaffian is a homogeneous
polynomial of degree $5$ in $7$ out of the $12$ moduli of ${\cal{M}}'$. An important pattern that emerges from the
results in~\cite{9,10} is that the smaller the dimension of ${\cal{M}}'$, the smaller the degree of the homogeneous polynomial.
A final comment is that the coefficient of proportionality has not been explicitly computed.

Now consider the simplified moduli space in which both ${\cal{M}}'$ and ${\cal{M}}_{{\cal{C}}'}$ are
one-dimensional and parameterized by the moduli $\psi$ and $\phi$ respectively. In this case, the
non-perturbative superpotential $W_{NP}$ in~(\ref{11}) takes the form
\begin{equation}
W_{NP}=A \psi^{p} e^{-\tau \phi},
\label{14}
\end{equation}
where $p$ is a small integer and $A$ is a coefficient of mass dimension $3-p$ in Planck units.
We now proceed to discuss the moduli dynamics in the presence of superpotential (\ref{14}).

\subsection*{Without Gravitation}

Again consider the dynamics of the moduli $\phi$ and $\psi$ in static flat spacetime. However, the complete superpotential is now given by
\begin{equation}
W=\lambda \phi^{2} \psi^{2} + A \psi^{p} e^{-\tau \phi}.
\label{15}
\end{equation}
The potential function describing the small instanton phase transition is
\begin{equation}
{\cal{V}}= |\partial_{\phi}W|^{2} + |\partial_{\psi}W|^{2},
\label{16}
\end{equation}
where
\begin{equation}
\partial_{\phi}W= 2 \lambda \phi \psi^{2}- \tau A \psi^{p} e^{-\tau \phi}, \quad
\partial_{\psi}W= 2 \lambda \phi^{2} \psi + p A \psi^{p-1} e^{-\tau \phi}.
\label{17}
\end{equation}
The minimum of this potential has two distinct types of separated loci. The first, specified by
\begin{equation}
\psi=0, \quad \phi \quad arbitrary,
\label{18}
\end{equation}
is the moduli space ${\cal{M}}_ {{\cal{C}}'}$ of ${\cal{C}}'$.
However, the moduli space ${\cal{M}}'$ is now replaced by $p-2$
isolated points located at
\begin{equation}
%Re \psi=0, \quad Im \psi= \pm \frac{2 \sqrt{2}}{\tau \sqrt{A} e^{2}}, \quad
\psi= \left(- \frac{2 \lambda p}{\tau^2 A e^p}\right)^{1/(p-2)}
e^{\frac{2 \pi i n}{p-2}}, \quad Re \phi=-\frac{4}{\tau}, \quad Im
\phi=0, \label{19}
\end{equation}
where $n=1, 2, \dots, p-2$.

To carry out an explicit calculation of the moduli dynamics, one
must specify the values of the parameters in superpotential
(\ref{15}). Without loss of generality, choose $\lambda=1$. For
specificity, we will take
\begin{equation}
\tau=400, \quad p=4, \quad A=\frac{1}{200e^{4}}.
\label{20}
\end{equation}
The choice for $\tau$ is consistent with the phenomenologically acceptable values of $R_{CY}$ and $\pi \rho$ given in~(\ref{13}). Similarly, in view of the discussion above, $p=4$ is a reasonable value for the degree of the Pfaffian polynomial. The choice of $A$ in~(\ref{20}) is less well-motivated. It is taken
because it enhances the distance between the local minimum and the ESP in moduli space, as can be seen from~(\ref{19}). For the parameters in~(\ref{20}), the local minima in~(\ref{19}) become
\begin{equation}
Re \psi=0, \quad Im \psi= \pm 0.1, \quad
Re \phi=-0.01, \quad Im \phi=0.
\label{21}
\end{equation}
Similar physical conclusions will be reached for a wide range of all three parameters.

With these choices, we compute the moduli dynamics  using initial conditions similar to those given in~(\ref{3}) and~(\ref{4}). The physical situation is identical to that described in the previous section. The moduli equations of motion are solved numerically on MATHEMATICA, with the results for $Re \phi$ and $Im \psi$ shown in Figure $6$.

\begin{figure}[htbp]
\center{\input{NoGrSol1.pstex_t}}
\caption{\label{pic16}{\footnotesize (C) and (B) represent the dynamics of $Re\phi$ and $Im\psi$ respectively for the case of a non-perturbative superpotential but no gravity. (A) gives the combined motion of these moduli.}}
\end{figure}

 Figure $6(C)$ indicates that the five-brane collides with the end of the interval but then, eventually, continues to oscillate without damping around that point. Now, however, there is a new phenomenon. Note that immediately after the collision the $Re \phi$ modulus gets ``hung up'' at the negative value corresponding to the local minima in~(\ref{21}). It lingers there for a considerable time, before rolling back through the origin and continuing its
oscillation around the small instanton point. Similar, but more dramatic, behaviour for $Im \psi$ is shown in Figure $6(B)$. After a few small oscillations around the origin of field space, the $Im \psi$ modulus rolls all the way out to the value $+0.1$ corresponding to one of the  local minima in~(\ref{21}). It stays in the vicinity of this minimum for some time, before rolling back through the origin and beginning an undamped oscillation around the small instanton. The choice of $+0.1$ rather than the other minimum at $-0.1$ is due to the initial conditions. These can be changed to select the other local minimum. The behaviour of both moduli is graphically presented in Figure $6(A)$, where the excursion of the moduli to the local minimum and back is clearly visible. The other two real moduli, $Im \phi$ and $Re \psi$, simply oscillate with undamped motion around the origin of field space, so we will not display them.

Comparing these results with the static flat case in Section $2$, we conclude that the non-perturbative superpotential has a dramatic effect on the dynamics of the moduli. Specifically, the moduli now evolve far from the origin of field space , seeking, and reaching, an isolated local minimum of the potential. However, since there is no mechanism for damping, the moduli roll through the isolated minimum, returning to oscillate around the small instanton. Were one to wait for a sufficiently long time, the moduli might once again briefly return to one of the two local minima, only to roll out again and oscillate around the origin of field space. The lack of any damping means that the overall oscillation amplitude persists. We conclude that, despite the existence of isolated local minima, the phase transition is never completed.

\subsection*{With Gravitation}

Now let the moduli $\phi$ and $\psi$ interact with gravity. The superpotential remains the same, namely
\begin{equation}
W=\lambda \phi^{2} \psi^{2} + A \psi^{p} e^{-\tau \phi}.
\label{22}
\end{equation}
Since the values of all moduli we consider, including the isolated local minima, are much smaller than unity, one can continue to choose the Kahler potential to be
\begin{equation}
K=|\phi|^{2}+|\psi|^{2}.
\label{23}
\end{equation}
Furthermore, since $\phi$ and $\psi$ are much smaller than one, the complicated potential function in~(\ref{7}) simplifies to
\begin{equation}
{\cal{V}}=|\partial_{\phi} W|^{2} + |\partial_{\psi} W|^{2}.
\label{24}
\end{equation}
That is, all terms that are higher order in $M_{P}^{-1}$ can be
ignored. The potential function is then identical to the
non-gravity case with identical minima; namely, the five-brane
branch~(\ref{18}) and the $p-2$ isolated minima given in~(\ref{19}).
As in the previous subsection, we choose $\lambda=1$ and take
\begin{equation}
\tau=400, \quad p=4, \quad A=\frac{1}{200e^{4}}.
\label{25}
\end{equation}
For these parameters, the two isolated minima in~(\ref{19}) are again
\begin{equation}
Re \psi=0, \quad Im \psi= \pm 0.1, \quad
Re \phi=-0.01, \quad Im \phi=0.
\label{26}
\end{equation}
Additionally, one must give an ansatz for the spacetime geometry. We will continue to assume an expanding FRW metric of the form given in~(\ref{9}).

Let us now compute the dynamics of the moduli $\phi$ and $\psi$, as well as the
spacetime expansion parameter $a$, using the same initial conditions as in Section $2$, namely, (\ref{3}),~(\ref{4}) and~(\ref{10}). The physical situation remains the same as previously. We use MATHEMATICA to numerically solve the equations of motion. The result for the scale parameter $a$ is similar to that shown in Figure $3$ and, hence, we won't present it here. The dynamics of $Re\phi$ and $Im\psi$ is given in Figure $7$.

\begin{figure}[htbp]
\center{\input{GrSol1.pstex_t}}
\caption{\label{pic11a}{\footnotesize (C) and (B) represent the
dynamics of $Re\phi$ and $Im\psi$ respectively for a
non-perturbative superpotential in an expanding FRW spacetime.
(A) gives the combined motion of these moduli.}}
\end{figure}

 Figure $7(C)$ indicates that the five-brane collides with the end of the interval. Immediately after the collision, the $Re\phi$ modulus gets ``hung up'' at the negative value corresponding to the local minima in~(\ref{26}). Now, however, the modulus is ``trapped'' near this value and does not return to the origin of field space. Rather, it oscillates with ever decreasing amplitude and frequency around $Re\phi=-0.01$, eventually coming to rest at this point. The behaviour of $Im\psi$ is similar. Figure $7(B)$ shows that after a few small oscillations around the origin of field space, the $Im\psi$ modulus rolls all the way out to the value $+0.1$ corresponding to one of the local minima of~(\ref{26}). Again, this local minimum, rather than $-0.1$, is selected by the initial conditions. However, unlike in the previous subsection, the modulus is now ``trapped'' near this minimum, oscillating around it with ever decreasing amplitude and frequency. It eventually comes to rest at $Im\psi=+0.1$.
The behaviour of both moduli is graphically presented in Figure $7(A)$. Here, the excursion of the moduli to the local minimum, and the phenomenon of their being trapped at that point, is clearly visible. The other two real moduli, $Im\phi$ and $Re\psi$, simply oscillate with decreasing amplitude and frequency around the origin of field space. Hence, we will not display them.

Comparing these results with those in the previous subsection, we see that the effect of the expanding FRW geometry on the dynamics of the $\phi$ and $\psi$ moduli is dramatic. Previously, without gravitation, these moduli evolved out to the local minimum of the non-perturbative superpotential. However, because of the absence of any mechanism for damping, the moduli eventually rolled out of the local minimum and back to the origin of field space. They oscillated forever with no loss of amplitude or frequency around the small instanton, occasionally, perhaps, temporarily revisiting one of the two local minima. When coupled to the expanding FRW spacetime, however, this behaviour is completely modified. The results in Figure$7$ graphically confirm that the moduli now evolve out to the local minimum and are trapped there by gravitational damping. The physical reason is clear. Immediately after collision, the moduli have sufficient ``energy'' to overcome a potential barrier and roll into a valley surrounding the local minimum. However, the energy is rapidly ``red-shifted'' by the expanding geometry. The result is that the moduli can no longer cross the potential barrier to return to the origin of field space.
The continued loss of energy to gravitational expansion causes the moduli to oscillate with smaller and smaller amplitude and frequency around the minimum, eventually coming to rest there. That is, they are gravitationally trapped at the isolated minimum.

Before drawing our final conclusions, we would like to comment on the effect of radiative damping on the moduli dynamics. As discussed in~\cite{13}, radiative damping occurs at an ESP, where the number of massless states and, hence,
the quantum mechanical amplitude to produce light particles is greatly
enhanced. This leads to radiative ``friction'' on the moduli dynamics, damping their motion. Thus far in this paper we have ignored this effect, considering only gravitational damping. Could the inclusion of radiative damping change our conclusions, perhaps trapping the moduli at the ESP? The answer is no. The
results reported in~\cite{13} show that although radiative damping is a strong effect, it requires the moduli to oscillate several times around the ESP before their amplitude is substantially reduced. However, as is clear from Figures $7(C)$ and $7(B)$, the transition of $\phi$ and $\psi$ from the vicinity of the ESP to that of the isolated minimum occurs immediately upon collision, within less than one oscillation (Note that the small amplitude oscillations of $Im\psi$ in Figure $7(B)$ all occur prior to the collision and not near the ESP). Furthermore, they never return to the ESP. It is safe to conclude, therefore, that radiative damping at the ESP simply has no opportunity to effect the moduli dynamics and can be safely ignored for the process under discussion in this paper. In fact, radiative damping helps enable our transition to the isolated minimum. The reason is that there is also an enhanced number of light states at the local minimum. Once trapped in its vicinity, radiative damping will work with gravitational damping to quickly bring the moduli to rest at the isolated minimum.

To conclude: In an expanding FRW spacetime, the collision of a five-brane with an end of the ${\bf S \rm}^{1}/{\bf Z \rm}_{2}$ interval passes through the small instanton point, described by the union of a vector bundle $V'$ and  a torsion free sheaf over the five-brane curve, and then continues to an isolated minimum of the non-perturbative superpotential where the vector bundle is smooth. Gravitational (and radiative) damping traps the moduli at the isolated minimum, thus completing the small instanton phase transition to the smooth vector bundle $V$. The moduli are
{\it not} attracted to the ESP in this process.

\section{Discussion:}

It has been demonstrated in this paper that not only is a
small instanton phase transition mathematically consistent, as shown in~\cite{11},
but it can also be realized dynamically. The results were presented
for a specific choice of parameters and initial conditions. However, we have
explored the dynamics of phase transitions for a range of parameters and initial conditions
and found that the transition is completed for many different values.
However, this is not always the case. Some choices do lead to trapping at the ESP, presumably because the energy is either too small to roll into the valley of an isolated minimum or so large that it rolls out again. Be this as it may, initial conditions leading to a completed phase transition are ubiquitous and demonstrate that a nonperturbative addition to the
potential can dramatically alter the results of~\cite{16,13,14,17,18}.

%%%%%%%%%%%%%%%%%%%%%%%%%%%%%%%%%%%%%%%%%%%%%%%%%%%%%%%%%%%%%%%%%%%%%%%%%%%%%%%%%%%%%%%%%%%%%%%%%%%%%%%%%%%%

\section{Acknowledgments:}

%%%%%%%%%%%%%%%%%%%%%%%%%%%%%%%%%%%%%%%%%%%%%%%%%%%%%%%%%%%%%%%%%%%%%%%%%%%%%%%%%%%%%%%%%%%%%%%%%%%%%%%%%%%%%%
We are grateful to Volker Braun for discussions of the intricacies of
MATHEMATICA. The work of Evgeny Buchbinder is supported by NSF grant PHY-0070928. The research of Alexander Borisov and Burt Ovrut is  supported in part by the Department of Physics and the Math/Physics Research Group at the University of
Pennsylvania under cooperative research agreement DE-FG02-95ER40893
with the U.~S.~Department of Energy and an NSF Focused Research Grant
DMS0139799 for ``The Geometry of Superstrings.''

%%%%%%%%%%%%%%%%%%%%%%%%%%%%%%%%%%%%%%%%%%%%%%%%%%%%%%%%%%%%%%%%%%%%%%%%%%%%%%%%%%%%%%%%%%%%%%%%%%%%%%%%%%%

\end{document}